# Overlay Accuracy Limitations of Soft Stamp UV Nanoimprint Lithography and Circumvention Strategies for Device Applications

P. J. Cegielski[a,b], J. Bolten[a], J. W. Kim[a], F. Schlachter[a], C. Nowak[a], T. Wahlbrink[a], A. L. Giesecke[a] and M. C. Lemme[a,b]

[a]AMO GmbH, Otto-Blumenthal-Str. 25, 52074 Aachen, Germany

[b]RWTH Aachen University, 52074 Aachen, Germany

**Abstract**

In this work multilevel pattering capabilities of Substrate Conformal Imprint Lithography (SCIL) have been explored. A mix & match approach combining the high throughput of nanoimprint lithography with the excellent overlay accuracy of electron beam lithography (EBL) has been exploited to fabricate nanoscale devices.

An EBL system has also been utilized as a benchmarking tool to measure both stamp distortions and alignment precision of this mix & match approach. By aligning the EBL system to 20 mm × 20 mm and 8 mm × 8 mm cells to compensate pattern distortions of order of 3 μm over 6 inch wafer area, overlay accuracy better than 1.2 μm has been demonstrated. This result can partially be attributed to the flexible SCIL stamp which compensates deformations caused by the presence of particles which would otherwise significantly reduce the alignment precision.

1. **Introduction**

When nanoimprint lithography (NIL) was introduced in 1996 by Chou [1] it was seen as a low cost alternative to conventional lithography. Since then, further research resulted in the development of numerous variations of NIL such us UV curable NIL (UV-NIL) [2], Step and Flash Imprint Lithography (S-FIL) [3] and soft stamp UV-NIL such as Substrate Conformal Imprint Lithography[4] (SCIL). SCIL, originally invented by Phillips research, uses a stamp copied from a pre-patterned master template typically fabricated using a conventional lithographic technique such as electron beam lithography (EBL), which is common for NIL methods. Its stamps consist of 3 layers: the first one, for pattern replication, is made of a high Young's modulus x polydimethylsiloxane (xPDMS) followed by a second buffer layer of soft PDMS and, as the third layer, a flexible 200 μm thick glass plate for support, which role is to reduce planar distortions of the stamp. These stamps enable wafer scale imprinting of 10 nm features at low process pressures due to their flexibility, which can follow a target substrate's topography, e.g. bow and warp of a standard silicon wafer [5]. At the same time this flexibility can lead to significant distortions of the stamp and replicated patterns, potentially limiting alignment precision of imprint to imprint or mix & match lithography using soft UV-NIL.

These limitations of the alignment accuracy achievable with SCIL have not been fully assessed until now. An alignment with overlay errors of 1 µm between two imprinted patterns by two different stamps was achieved by correction of a systematic error introduced by lateral displacement of the working stamp occurring during imprinting [6]. Such limited overlay precision reduces the usefulness of SCIL for micro- and nanoscale devices, despite its general potential for high resolution large area patterning. As a consequence, SCIL is currently mostly limited to large area patterning of periodic structures. The overlay precision of 1 µm demonstrated in the referenced work is thought to be limited by the mechanical precision of the mask aligner used for these experiments, which is approximately 1 µm [6]. In order to obtain a more complete picture of SCIL capabilities and limitations for multilevel patterning, we utilized mix & match lithography using SCIL and EBL. This approach ensures that alignment errors originate nearly exclusively from the SCIL process because EBL can achieve an alignment precision better than 5 nm, if the set of EBL alignment marks is well defined and the substrate is conductive [7]. Hence, errors introduced by the EBL system can be neglected compared to errors introduced by other process steps. Such evaluation is of great importance for broadening the application range of SCIL. Hence, in this work we explore those limitations and present one potential application for SCIL/EBL mix & match lithography in nanophotonics.

2. Experiment

Many EBL systems allow high precision automated marker detection at predefined positions. This makes them ideally suited to acquire marker position data, which can be translated into a distortion map of patterns defined by a SCIL process. Furthermore, the marker position data can be compared with the expected marker positions and used to calculate both global and local rotation, scale and keystone corrections. These corrections enable the definition of patterns which partially compensate the positioning errors introduced by SCIL.

In order to evaluate the distortion of imprinted patterns, as well as EBL to SCIL alignment precision, a special layout was designed containing twelve 20 mm × 20 mm cells covering a full 6'' wafer area (Fig. S1 of Supporting Information). Each cell contained four 10 µm × 10 µm rectangular EBL alignment markers and nine evenly spaced Vernier scales for both X and Y axes, which allow reading the overlay error with 100 nm precision (Fig. 1). In addition, each 20 mm × 20 mm field contained four 8 mm × 8 mm sub-cells with individual sets of 4 EBL alignment markers and 9 Vernier scales. These sub-cells allowed to investigate whether a reduction of the cell size and hence a more localized alignment and compensation of local distortions of the pattern influences the alignment precision.

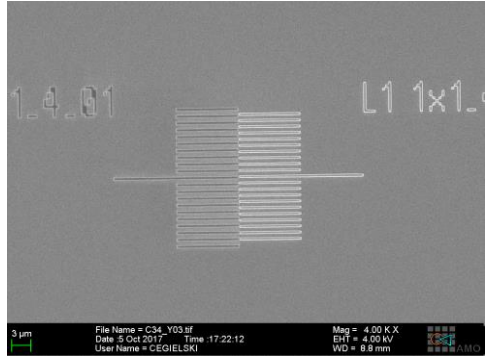

Fig. 1 Vernier scale used to evaluate the alignment precision in Y direction. The left side was patterned by EBL and the right side by SCIL.

The fabrication of an imprint working stamp started with preparation of a Si master template. For this purpose a Si wafer was coated with PMMA resist, which was exposed with an e-beam and subsequently developed using a 7:3 mixture of IPA and water. The pattern was then transferred by inductively coupled plasma reactive ion etching (ICP RIE) with $SF_6$ and $C_4F_8$ gases 100 nm deep into the master wafer. After resist removal, the obtained Si master was coated with antiadhesive $C_4F_8$ polymer and ~1.5 μm of xPDMS. Afterwards xPDMS was joined with a support glass plate (Schott AF32) by a PDMS buffer layer. The resulting three layer SCIL stamp was cured for 48 h in 50 °C, followed by a release from the master.

EBL alignment marks and the right side of the Vernier scales (Fig. 1) were defined in a single SCIL process. As the alignment precision of the EBL is highly depending on the quality of the markers, special care has been taken to ensure the definition of high quality and high contrast markers in the alignment test wafer. The marker features had to be transferred ~500 nm deep into silicon. Therefore, SCIL was utilized to pattern a 40 nm thick highly selective Cr hard mask by a bi-layer resist lift-off process: AMONIL imprint resist was spun on a 50 nm thick PMMA layer. Imprinting was performed using a SCIL imprint tool based on the MA8 mask aligner series (SÜSS Microtec GmbH) with pressure of 30 mbar at a speed of the contact front of 1 s per groove. After imprinting and etching of the AMONIL residual layer, this sacrificial PMMA was etched in $O_2$ plasma exposing the surface of the Si wafer. Next, Cr was evaporated and the wafer was placed in acetone, which dissolved PMMA allowing a lift-off of the otherwise insoluble AMONIL and Cr layers. Afterwards the pattern was transferred by $SF_6$ and $C_4F_8$ ICP RIE and the residual Cr mask was removed by wet etching using a commercially available mixture of perchloric acid and ceric ammonium nitrate. In a next step the second half of Vernier scales was defined in HSQ resist by EBL. The marker structures defined in the SCIL master and replicated by the SCIL process discussed above were used to align both sets of Vernier scales. The alignment precision was then measured by analyzing scanning electron micrographs of the Vernier scales.

3. Results and discussion

The EBL system successfully aligned to every design cell by the automated marker search routine. At the same time it provided exact data regarding the absolute and relative positions of the alignment markers defined by NIL. A few markers were not defined by SCIL due to imprint defects leading to uncharacteristically large alignment errors in some of the cells, which have therefore been excluded from further considerations. The data set was used to calculate the difference between the expected positions and observed positions of the markers with respect to a reference marker located close to the wafer center including corrections for the wafer rotation (Fig. 2). The markers displacements in both X and Y directions evidence a global compression of the imprinted pattern with a slope of 0.01µm/mm (Fig. S2 of Supporting Information), and reached 3 µm in the outermost design cells. Taking into the account that the stamp was cured at 50 °C and then cooled to room temperature it is clear that the imprinted pattern was scaled down with an effective thermal expansion coefficient ($\alpha$) of 0.33 ppm/K (in agreement with previously reported value [8]), which is two times smaller than the actual difference between $\alpha$ of Si master wafer (2.6 ppm/K) and AF32 stamp support glass (3.2 ppm/K). The marker displacement reached 3 µm and was 1.65 µm on average therefore it could not be explained only by the thermal mismatch of Si and AF32 glass, which could distort the pattern only by 600 nm over 60 mm distance. Hence, contributions of the imprint process must be accounted for.

20 mm × 20 mm cells dimensions differed from the nominal values ($\Delta X$ and $\Delta Y$) by max ± 3µm. The average $\Delta X$ was close to 0 with σ of 1.5 µm while in Y the cells were mostly stretched (average $\Delta Y$ of 0.54 µm) and the error was more systematic indicated by lower σ = 1 µm. In 8 mm × 8 mm cells the max $\Delta X$ and $\Delta Y$ were similar to maximal values observed in the larger cells; however the standard deviation was smaller by one third meaning those deformations have mostly local nature.

The pattern distortions can be divided into 3 categories depending on their source: i) displacement due to the thermal contraction of the stamp after curing, ii) stretching of cells in the Y direction due to imprinting and iii) random, local cell deformations in X and Y. Marker displacement was also measured on the Si master wafer, resulting in an average value of 115 nm (Fig. S3). This confirms that discussed displacements are not present in the initial master wafer and mainly originate from the imprinting process.

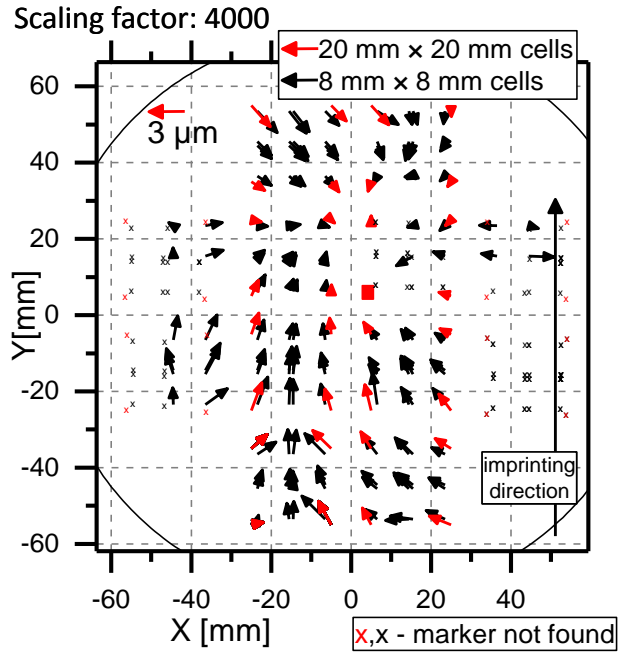

Fig. 2 Displacement of EBL alignment markers of 20 mm × 20 mm (red) and 8 mm × 8 mm (black) cells defined by SCIL with respect to the reference marker (red box). The displacement vector tail marks the expected position and the arrow head marks the observed position. The vectors are scaled up by a factor of 4000 compared to the wafer position (indicated by the 3 μm scale arrow). The left most and right most cells were not fully imprinted because the SCIL tool could not provide full contact of the stamp with the wafer. 6''wafer is outlined with a black line.

On the overlay error map of a cell located in the rightmost upper side of the wafer obtained by extracting the values from the Vernier scales (Fig. 3, full mapping for all 12 cells is available in Fig S4 of the supporting information) one can see that alignment precision was different for each 8 mm × 8 mm sub cell. Sub cells located in the left column showed a higher alignment error which was directly related to higher cell deformations than in the right column sub cells. On a scale of the entire wafer it can be seen that the alignment error is directly correlated to the average of |ΔX| and |ΔY| in each cell. The alignment error σ was growing by ~120 nm for every μm of average cell deformation (Fig. S5 of supporting information). No correlation was found when deformations in only X or Y directions were taken into account. Whether the cell was mostly stretched or compressed was not correlated with the alignment error as well.

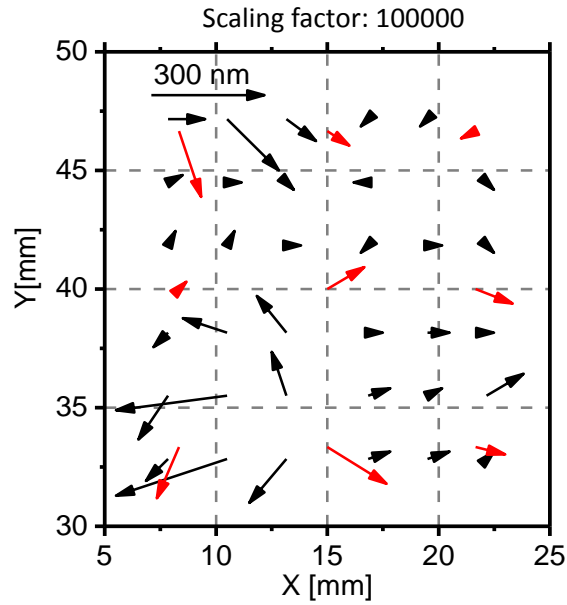

Fig. 3 Map of alignment error extracted from the Vernier scales of the right most upper 20 mm × 20 mm cell (red) including 4 8 mm × 8 mm sub cells (black). Arrows (scaled 100 000 times indicated by 300 nm scale arrow) illustrate the direction and magnitude of misplacement of pattern defined by EBL relative to pattern defined by NIL. A map of the full wafer is available in the supporting information in the figure S4.

The statistical analysis of the measured overlay error obtained from the cells, in which alignment markers were found by EBL system (Fig. 4(a)) shows a standard deviation of $\sigma$ = 361 nm and $\sigma$ = 381 nm for X and Y directions for 20 mm × 20 mm cells, with a small offset of 120 nm in X direction and 43 nm in Y direction. Reducing the cell size to 8 mm × 8 mm resulted in an overlay error $\sigma$ = 376 nm in X and $\sigma$ = 322 nm in Y directions, with only slight offsets of 6 nm in X and -64 nm in Y direction (Fig. 4 (b)). The 3$\sigma$ value, which corresponds to probability of 99.7 %, is in the range of 1.2 μm.

The alignment precision was similar in both X and Y directions and also across the wafer indicating that the systematic pattern distortions, caused by thermal effects and those related to the imprint direction, were well compensated by the EBL system. Therefore the errors that could not be corrected must have their source in local and nonlinear pattern distortions. Particle contamination was initially expected to be the cause of those deformations. However, it was observed that errors red out from Vernier scales which were located directly next to particles (Fig. S6 of Supporting Information), which locally deformed the stamp preventing contact with the substrate, were often very low. Hence, the flexible SCIL stamp is capable of compensating the particle contamination to maintain the alignment precision.

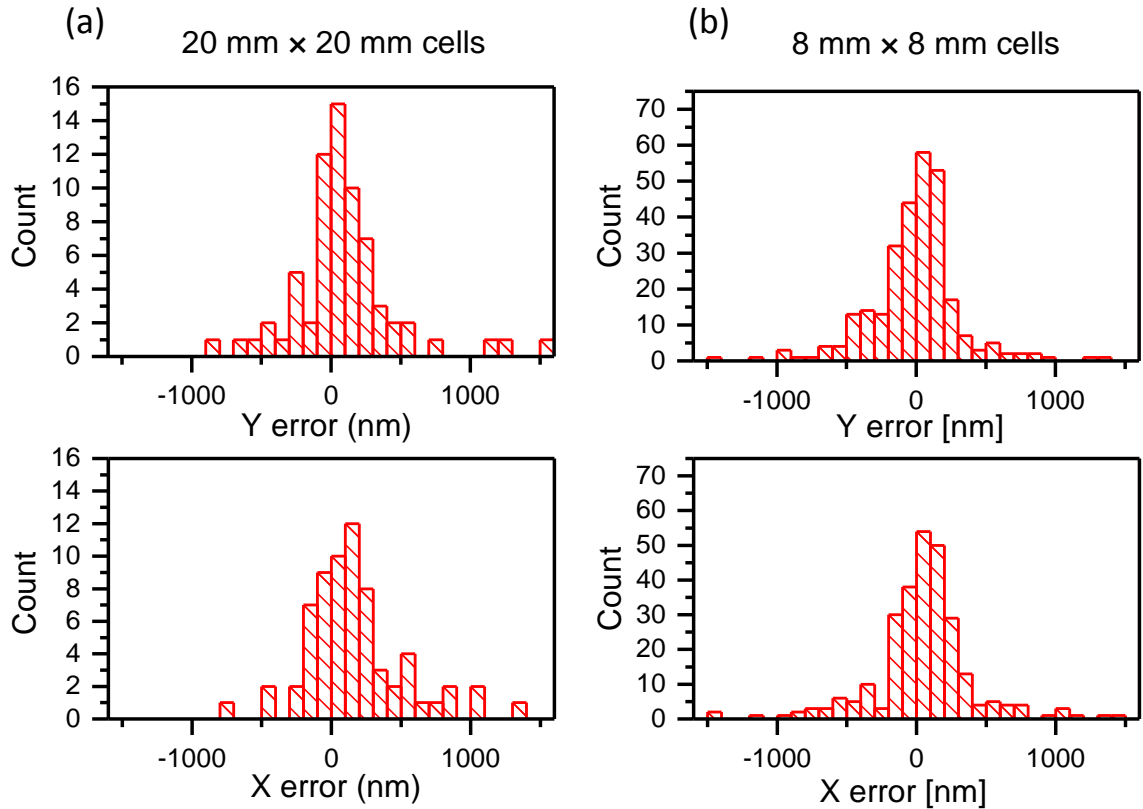

Fig. 4 Statistical distribution of alignment error measured with Vernier scales in (a) 20 mm × 20 mm cells (b) 8 mm × 8 mm.

## 4. Device application

Nanophotonic devices based on planar waveguides often require patterning with ~100 nm resolution to ensure critical dimension control, which is important for correct realization of optical functions. In many applications the waveguide layer, which is patterned first, is also the only layer requiring high patterning precision and additional layers such as metallization (e.g. electrodes [9]) have relaxed requirements. Both hold true for the demonstrator presented here. It consists of a silicon nitride double slot waveguide with minimum CD features of ~100 nm (width of the slots) and two large Al electrodes with a 4 μm wide gap, which can be used to control optical properties of an organic functional material deposited on the waveguide. The alignment tolerance for waveguides and electrode layers is ±1 μm, which ensures that light propagating in the waveguide cannot reach the electrode. The waveguides were patterned in a 250 nm thick $Si_3N_4$ layer on 2.2 μm thermal $SiO_2$ on a 6'' Si handle wafer by using a 20 nm Cr hard mask patterned by lift-off with double AMONIL/PMMA resist, as described above. Aluminum electrodes were then patterned by liftoff utilizing a PMMA/MMA bi-layer resist stack patterned by EBL (Fig. 5). The alignment precision better than 0.8 μm was achieved by using 20 mm × 20 mm cells, which was within the devices tolerance. However for some cells and devices much better results were achieved; alignment error of order of 50 nm could be measured as can be seen in Fig. 5.

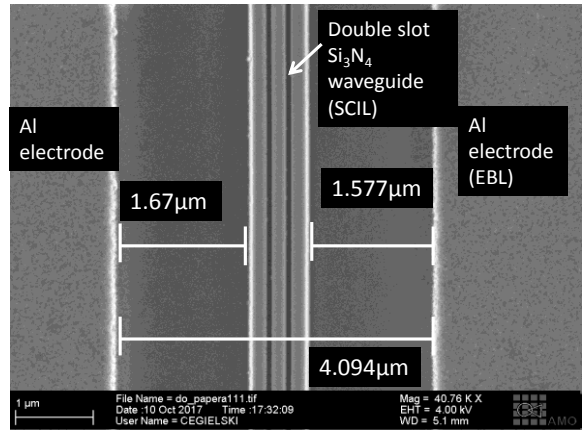

Fig. 5 Nanophotonic device with double slot silicon nitride (Si$_3$N$_4$) waveguides fabricated by SCIL and EBL mix & match lithography.

5. Conclusions

Large global distortions of the imprinted pattern were quantified and reached values as high as 3 μm caused by thermal expansion coefficient mismatch between the Si master and the stamp support glass and by stamp deformations during imprinting. Such distortions are strongly limiting the possibility of full wafer alignment without making corrections, such as those performed by EBL systems, which can efficiently compensate such systematic errors. Local distortions of the pattern evaluated by EBL exposure yielded 3σ of 1.2 μm. Since the systematic errors were corrected by the EBL system and EBL is capable of achieving 5 nm alignment precision the obtained 3σ value arises exclusively from nonlinear distortions of the imprinted pattern. Hence, 3σ of 1.2 μm is the practical limit of alignment precision of SCIL, which can be improved only by a careful optimization of the imprinting process.

If the alignment tolerance of a specific application allows SCIL can be used for high resolution patterning of the first layer to reduce the total cost per device, once the design is finalized so that fabrication of the imprint master is justified. Defining a second layer by EBL as a flexible and fast prototyping technique can then significantly speed up development cycles. This approach is particularly useful in nanophotonics prototyping, where a single waveguide design can be used as a basic building block which can be extended to different functional devices by adding further device features using other lithographic techniques such as EBL.

Acknowledgment

European Union's Horizon 2020 research and innovation programme under the Marie Sklodowska-Curie grant agreement No. 643238 (Synchronics) and PPP No. 688166 (Plasmofab).

References

[1]    S.Y. Chou, P.R. Krauss, P.J. Renstrom, Science (80-. ). 272 (1996) 85–87.

[2]  J. Haisma, J. Vac. Sci. Technol. B Microelectron. Nanom. Struct. 14 (1996) 4124.

[3]  M. Colburn, S.C. Johnson, M.D. Stewart, S. Damle, T.C. Bailey, B. Choi, M. Wedlake, T.B. Michaelson, S. V. Sreenivasan, J.G. Ekerdt, C.G. Willson, in:, Y. Vladimirsky (Ed.), SPIE Proc., 1999, p. 379.

[4]  M. Verschuuren, Substrate Conformal Imprint Lithography for Nanophotonics, 2009.

[5]  R. Ji, M. Hornung, M.A. Verschuuren, R. van de Laar, J. van Eekelen, U. Plachetka, M. Moeller, C. Moormann, Microelectron. Eng. 87 (2010) 963–967.

[6]  R. Fader, M. Rommel, A. Bauer, M. Rumler, L. Frey, M. Antonius Verschuuren, R. van de Laar, R. Ji, U. Schömbs, J. Vac. Sci. Technol. B, Nanotechnol. Microelectron. Mater. Process. Meas. Phenom. 31 (2013) 06FB02.

[7]  J. Bolten, N. Koo, T. Wahlbrink, H. Kurz, Microelectron. Eng. 110 (2013) 224–228.

[8]  R. Fader, M. Förthner, M. Rumler, M. Rommel, A.J. Bauer, L. Frey, M.A. Verschuuren, J. Butschke, M. Irmscher, E. Storace, R. Ji, U. Schömbs, in:, NNT 2014, 13th Int. Conf. Nanoimprint Nanoimprint Technol., Kyoto, Japan, 2014.

[9]  F. Qiu, A.M. Spring, D. Maeda, M. Ozawa, K. Odoi, A. Otomo, I. Aoki, S. Yokoyama, Sci. Rep. 5 (2015) 8561.

# Supplementary information

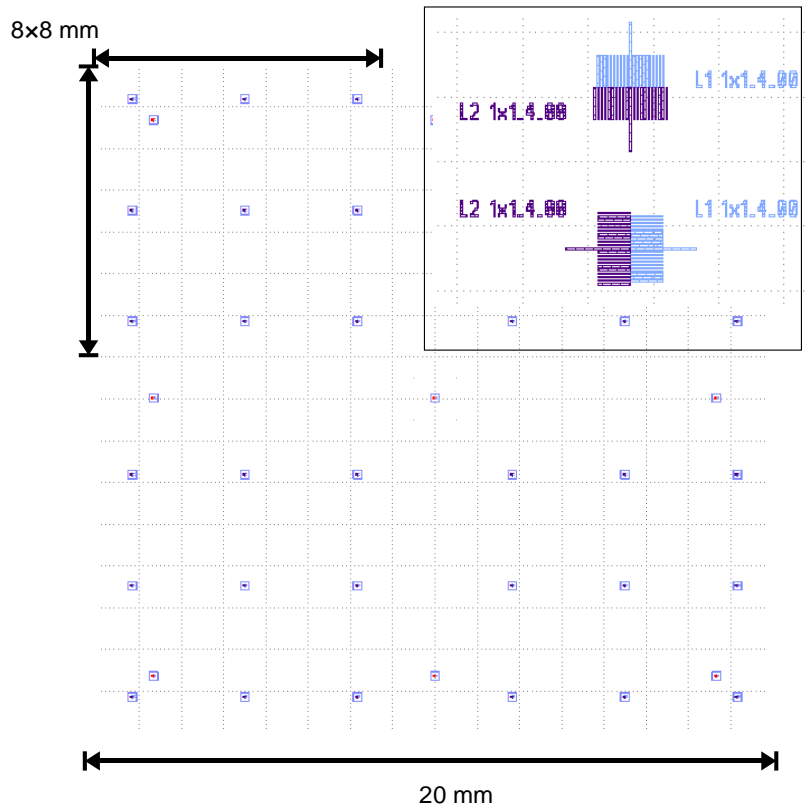

S 1  Layout of a 20 mm × 20 mm cell including 4 8 mm ×8 mm sub cells. Each cell contains 9 Vernier scales (Inset) for measuring alignment error in X and Y directions with 100 nm precision

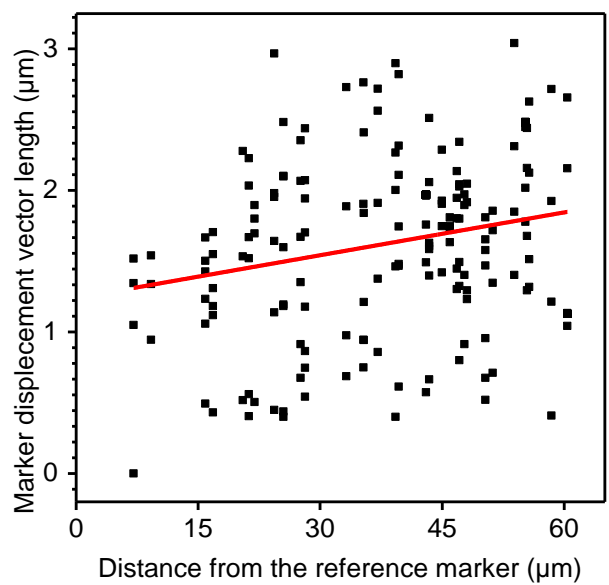

S 2 Marker displacement vector length vs. the distance from the wafer center. The slope of 0.01 µm/mm corresponds to a negative thermal magnification (0.33 ppm/K) of the pattern due to thermal mismatch between AF 32 stamp support glass and Si master wafer.

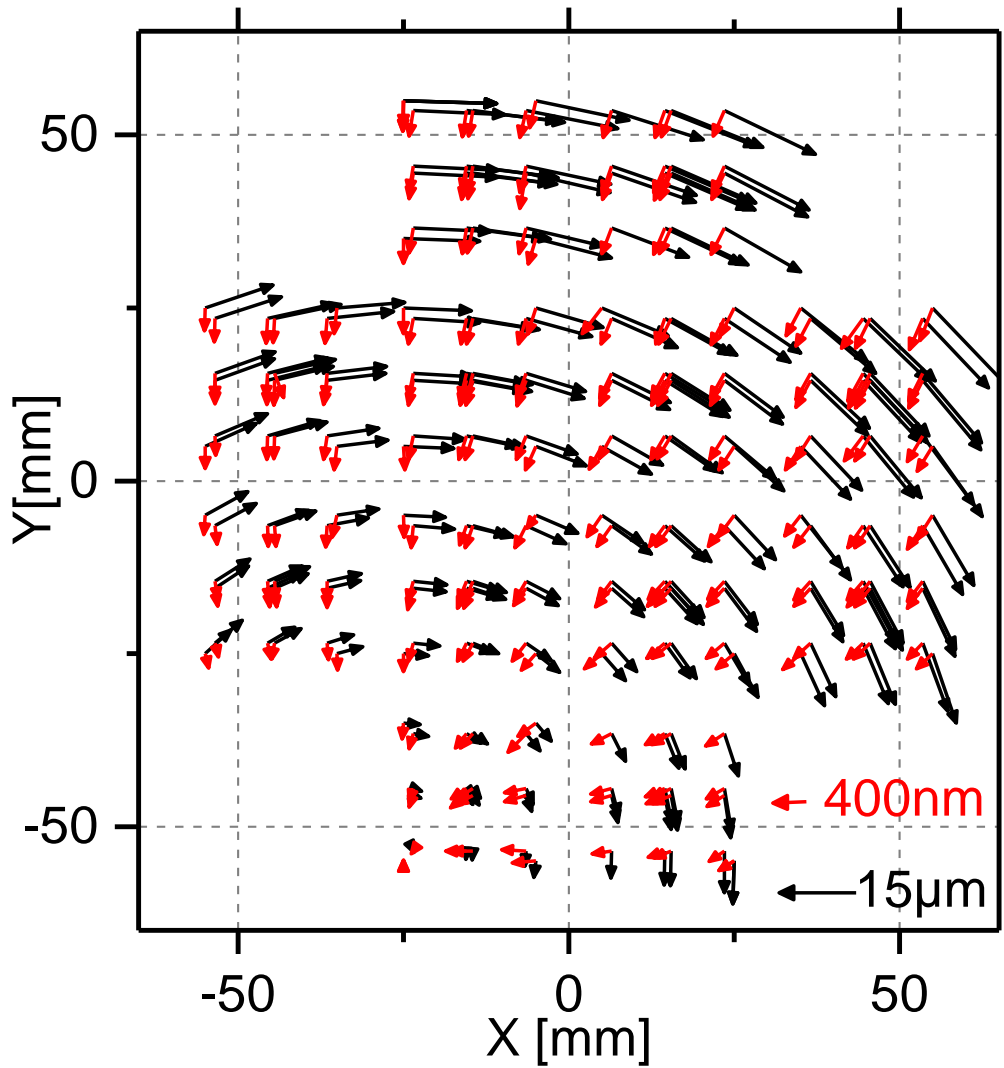

S 3 Black: Measurement of positions of alignment markers on Si master with respect to a marker located in the bottom left corner. Initially only the rotation originating from placing the wafer in the wafer holder is visible. Red: Measured positions after subtracting rotation. The mean marker misplacement in X and Y is -167 and - 344 nm while standard deviation is 115 nm for both axes. Since mean values are greater than the standard deviation it is clear that the rotational error was not fully corrected and the real marker displacement is smaller.

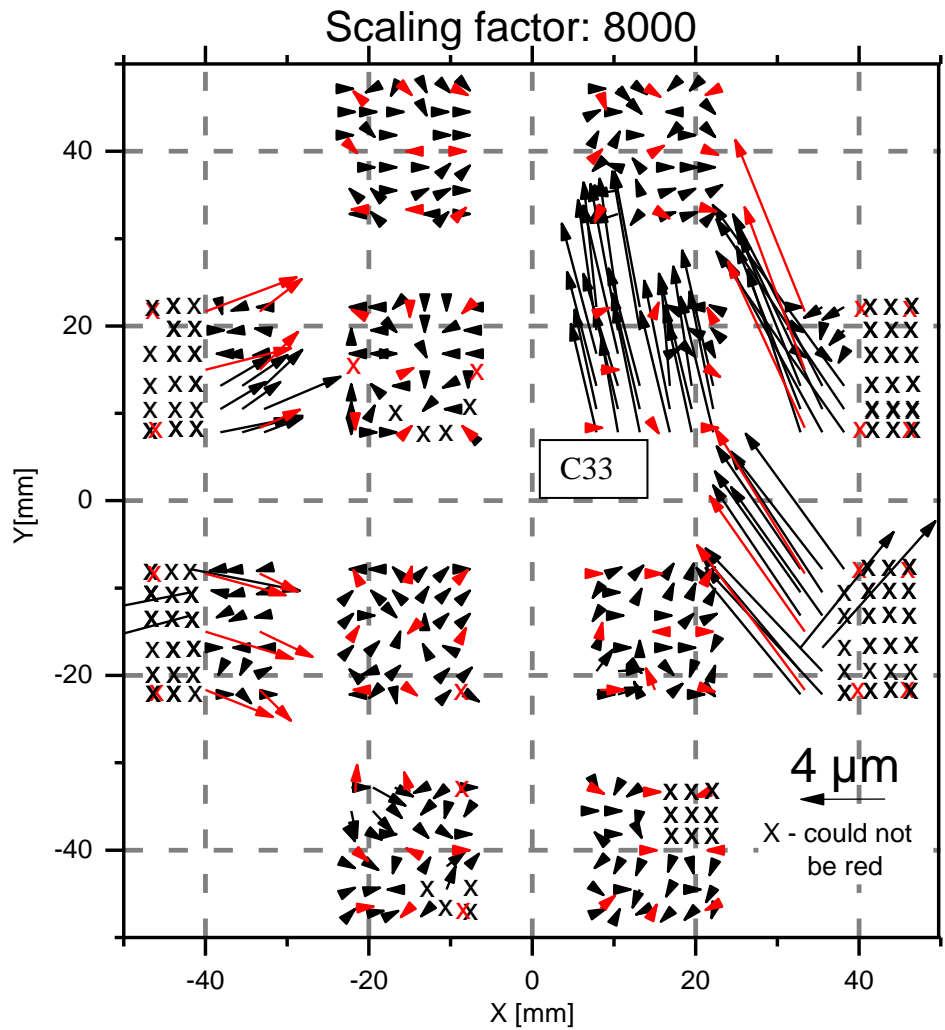

S 4 Map of the overlay error of EBL structures aligned to pattern defined by SCIL using 20 mm × 20 mm alignment cells (red) and 8 mm by 8 mm sub cells (black). In the cell marked C33 alignment markers were not found by EBL system in each 3 out of 4 sub cells, therefore distortions were not properly compensated.

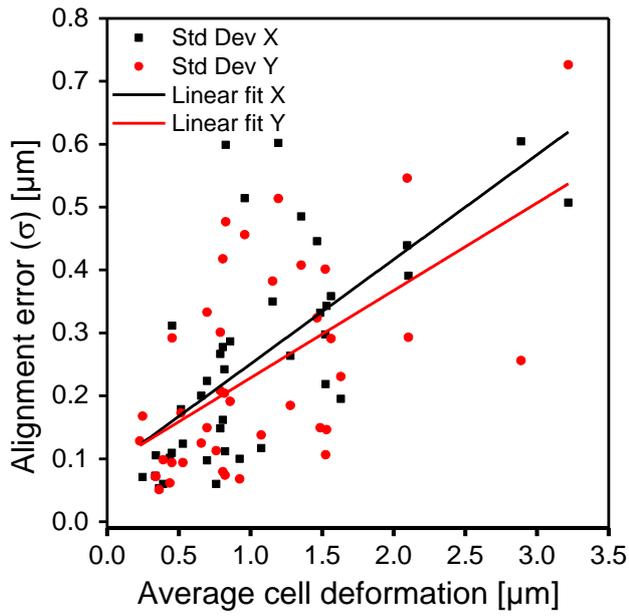

S 5  Standard deviation (σ) of the alignment error in 8 mm × 8 mm cells vs. the average of absolute values of cell deformation in both X and Y directions. Solid lines are linear fits with slope of 166 nm and 132 nm per 1 μm.

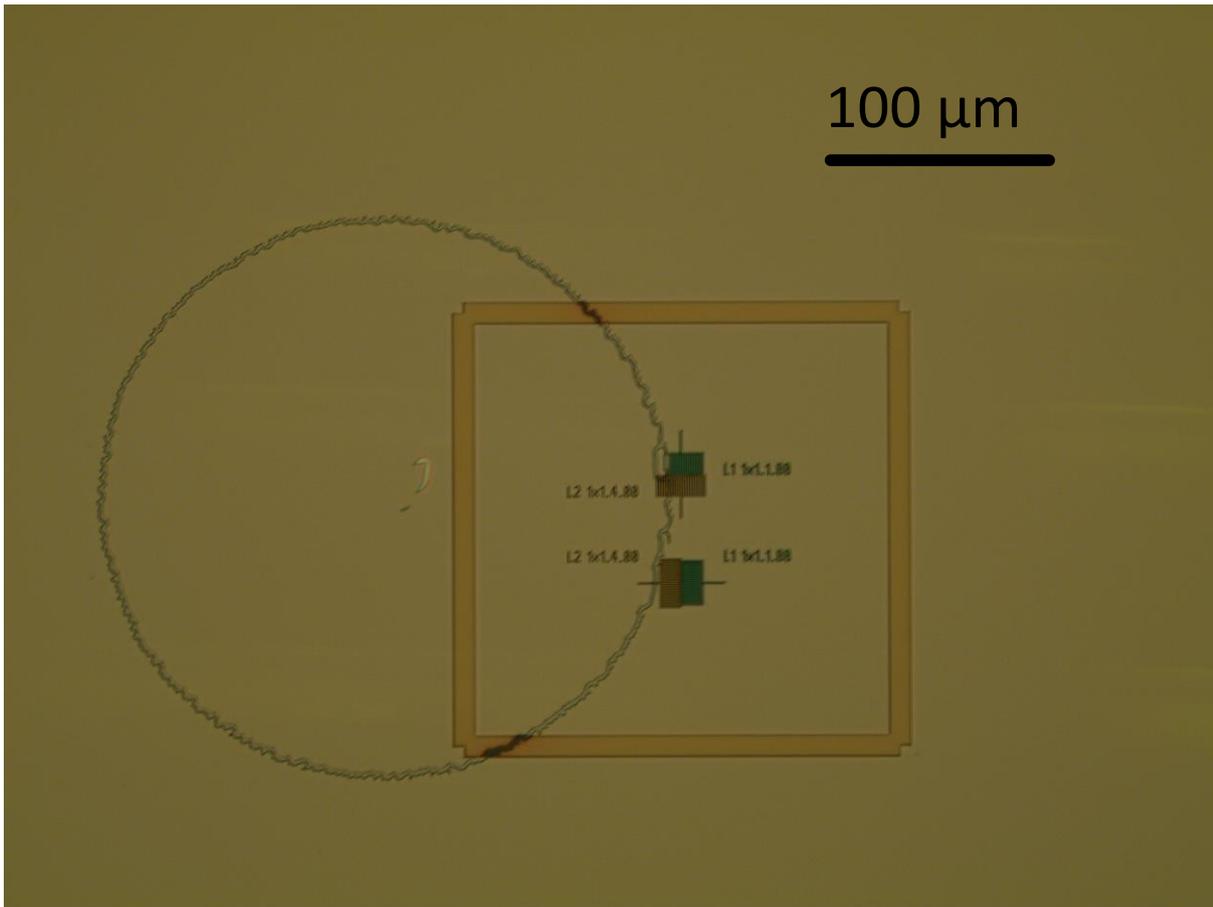

S 6 A Vernier structure imprinted next to a particle. Around the particle the stamp could not contact the wafer surface leaving a bubble of ~ 100 μm radius. Despite that the alignment error measured by this Vernier scale was less than 100 nm in X and less than 200 nm in Y directions.